\documentclass[fleqn,10pt]{wlscirep}
\usepackage[utf8]{inputenc}
\usepackage[T1]{fontenc}
\usepackage{bm}
\usepackage{lscape}
\usepackage{marvosym}
\usepackage{color}
\usepackage[FIGTOPCAP,nooneline]{subfigure}

\title{A comprehensive survey on quantum computer usage: How many qubits are employed for what purposes?}

\author[1,*]{Tsubasa Ichikawa}
\author[1]{Hideaki Hakoshima}
\author[2]{Koji Inui}
\author[1]{Kosuke Ito}
\author[3]{Ryo Matsuda}
\author[1,3]{Kosuke Mitarai}
\author[1]{Koichi Miyamoto}
\author[1]{Wataru Mizukami}
\author[2,4,5]{Kaoru Mizuta}
\author[1,2]{Toshio Mori}
\author[3]{Yuichiro Nakano}
\author[3]{Akimoto Nakayama}
\author[1]{Ken N. Okada}
\author[1,6,7]{Takanori Sugimoto}
\author[3,8]{Souichi Takahira}
\author[1,9]{Nayuta Takemori}
\author[1]{Satoyuki Tsukano}
\author[1,6]{Hiroshi Ueda}
\author[3]{Ryo Watanabe}
\author[1]{Yuichiro Yoshida}
\author[1,2,3]{Keisuke Fujii}

\affil[1]{Center for Quantum Information and Quantum Biology, Osaka University, 1-2 Machikaneyama, Toyonaka, Osaka 560-0043, Japan.}
\affil[2]{Center for Quantum Computing, RIKEN, Hirosawa 2-1, Wako Saitama 351-0198, Japan.}
\affil[3]{Graduate School of Engineering Science, Osaka University, 1-3 Machikaneyama, Toyonaka, Osaka 560-8531, Japan.}
\affil[4]{Department of Applied Physics, Graduate School of Engineering, The University of Tokyo, 7-3-1 Hongo, Bunkyo-ku, Tokyo 113-8656, Japan.}
\affil[5]{Photon Science Center, Graduate School of Engineering, The University of Tokyo, 7-3-1 Hongo, Bunkyo-ku, Tokyo 113-8656, Japan.}
\affil[6]{Computational Materials Science Research Team, RIKEN Center for Computational Science (R-CCS), Kobe, Hyogo 650-0047, Japan.}
\affil[7]{Advanced Science Research Center, Japan Atomic Energy Agency, Tokai, Ibaraki 319-1195, Japan.}
\affil[8]{Faculty of Information Engineering, Meijo University, 1-501 Shiogamaguchi, Tempaku-ku, Nagoya 468-8502, Japan.}
\affil[9]{Center for Emergent Matter Science, RIKEN, Wako, Saitama 351-0198, Japan}

\affil[*]{e-mail: ichikawa.tsubasa.qiqb@osaka-u.ac.jp}

\begin{abstract}
Quantum computers (QCs), which work based on the law of quantum mechanics, are expected to be faster than classical computers in several computational tasks such as prime factoring and simulation of quantum many-body systems.
In the last decade, research and development of QCs have rapidly advanced.
Now hundreds of physical qubits are at our disposal, and one can find several remarkable experiments actually outperforming the classical computer in a specific computational task.
On the other hand, it is unclear what the typical usages of the QCs are.
Here we conduct an extensive survey on the papers that are posted in the quant-ph section in arXiv and claim to have used QCs in their abstracts. 
To understand the current situation of the research and development of the QCs, we evaluated the descriptive statistics about the papers, including the number of qubits employed, QPU vendors, application domains and so on.
Our survey shows that the annual number of publications is increasing, and the typical number of qubits employed is about six to ten, growing along with the increase in the quantum volume (QV).
Most of the preprints are devoted to applications such as quantum machine learning, condensed matter physics, and quantum chemistry, while quantum error correction and quantum noise mitigation use more qubits than the other topics.
These imply that the increase in QV is fundamentally relevant, and more experiments for quantum error correction, and noise mitigation using shallow circuits with more qubits will take place.

\end{abstract}
\begin{document}

\flushbottom
\maketitle

\thispagestyle{empty}

\noindent \textbf{Key points:} \\
$\bullet$ The annual number of manuscripts actually using quantum computers is growing and seems to be saturated in the last two years.\\
$\bullet$ The average number of the qubits employed has been growing along with the increase in quantum volume and reached 10.5, whereas the median attained 6. \\
$\bullet$ The preprints on quantum error correction, quantum noise mitigation, and system and software development tend to employ more qubits than the other topics. \\
$\bullet$ Over half of the total number of preprints are for the applications such as condensed matter physics, quantum chemistry, and quantum machine learning, to name a few.



\section*{Introduction}
The last decade has witnessed great progress in the research and development of quantum computers (QCs). 
Starting with the opration of qubits and gates on them below the error threshold of the surface code\cite{Barends2014}, we have witnessed the cloud release of accessible quantum processing units (QPUs) followed by the rapid increase in the number of qubits in QPUs\cite{ezratty2023heading}, and the demonstrations of quantum computational supremacy\cite{Arute2019-rf,doi:10.1126/science.abe8770,PhysRevLett.127.180501,ZHU2022240,Madsen2022-ki}.
Now we can manipulate over 100 qubits to simulate the ground states of the quantum spin chains\cite{PhysRevResearch.5.013183} and dynamics of the two-dimensional transverse-field Ising model~\cite{kim2023evidence}.

The above perspective appropriately captures the recent progress, but we should remember that it consists of outstanding experiments, which are part of many others; and thereby we may overlook how typical usages of QCs have been.
Such oversight, if exists, could lead to an insufficient understanding of the present situation in the research and development of QCs and may result in possible obstacles to long-term research design.
In other words, we need an alternative perspective based on typical usages as well as the standard one based on monumental experiments in order to properly construe the present situation.

In this review article, we conduct a bibliographic survey to describe the typical usage of QCs.
For the preprints posted in the quant-ph section of arXiv from January 1 of 2016 to November 10 of 2022, we evaluate descriptive statistics of the annual number of submissions and the number of employed qubits with the classifications by QPU vendors, authors affiliations, and topics of the preprints. 
Through these analyses, we find the following:
\begin{enumerate}
    \item The annual number of manuscripts using quantum computers is growing mainly due to the increase in the submissions from the authors not affiliated with the QC manufacturers (non-vendor users), and this growth seems to be saturated in the last two years.
    The use of IBM devices dominates over that of the other companies\rq~devices.
    \item The average number of the qubits employed has been growing along with the increase in quantum volume (QV) \cite{PhysRevA.100.032328} and reached 10.5, whereas the median has attained 6.
    \item The preprints on quantum error correction, quantum noise mitigation, and system and software development tend to employ more qubits than the other topics.
    In particular, the rapid increase in the number of employed qubits is observed in works on the quantum error correction conducted by the authors affiliated with the QC manufacturers (vendor users).
    The experiments employing many qubits use shallow circuits implementable with NISQ (Noisy Intermediate-Scale Quantum)  devices\cite{Preskill2018quantumcomputingin}.
    \item Over half of the total number of the preprints are for the applications such as condensed matter physics, quantum chemistry, quantum machine learning, optimization by quantum approximate optimization algorithm (QAOA), use of primitive algorithms, simulations of quantum mechanics, and high energy physics.
    Non-vendor users submitted most of the preprints in those applications.

\end{enumerate}

Our findings quantitatively lead us to the following outlook: 
Having demonstrated the feasibility of large-scale experiments of quantum error correction and quantum simulation of condensed matter with the use of the NISQ devices, we will observe the increase in the number of such experiments with more qubits employed.
In parallel, two directions are relevant for an increase in the number of larger experiments conducted by the non-vendor users aiming at the applications of QCs.
One is that the skillful users design shallow circuits suitable for the topics that they deal with.
The other is that the vendors increase the number of qubits on which we can reliably implement the quantum circuits.
The number of such qubits could be measured with QV, which is a metric for executing relatively deep quantum circuits 
independent of hardware connectivity.
The increase in QV is also desirable for majority of users to employ more qubits in their applications leading to increasing average number of qubits employed.

\section*{Material}
In this survey, we grasp the present situation in the academic usage of QCs mainly through descriptive statistics of the preprints uploaded in the subsection quant-ph of the arXiv repository.
To this end, we collected the preprints that  actually use QCs as many as possible.
Here we list our procedure for the data collection process.

To select the preprints potentially using QCs from arXiv, we first chose the following five companies (hereafter called vendors) as the targets to analyze how their QCs are employed: Google, IBM, IonQ, Quantinuum (Honeywell before the amalgamation with Cambridge Quantum Computing), and Rigetti. This choice was based on the fact that these companies have been listed in the US stock markets (New York Stock Exchange or NASDAQ) and launched the research and development of QCs at the early stage.
Microsoft is excluded from our analysis since no QCs are launched.
Second, we set the time span to be analyzed (the survey period), January 1 of 2016, which is the year that IBM announced the cloud service of QC\cite{Mandelbaum_2021}, through November 10 of 2022.

\begin{table}[t]
    \centering
    \begin{tabular}[t]{c|l}
    \hline
        Code & Topic 
        \\ \hline
    \hline
        A & Fundamentals of physics\\
        ~ & and quantum information\\
        B & Tomography, noise characterization,\\
        ~ & quantum control (including pulse 
        \\ 
        ~ &  optimization), gate benchmarking
        \\
        C & System and software development\\
        ~ & for quantum computers
        \\ 
        D & Quantum error correction 
        \\ 
        E & Quantum noise mitigation 
        \\ 
        F & Applications 
        \\ \hline
    \end{tabular} 
    $\qquad$
    \begin{tabular}[t]{c|l}
    \hline
        Code & Subtopic in F: Applications 
        \\ \hline
    \hline
        F1 & Quantum machine learning\\
        F2 & Condensed matter physics  
        \\ 
        F3 & Quantum chemistry
        \\ 
        F4 & Finance 
        \\ 
        F5 & Optimization (QAOA) 
        \\ 
        F6 & Linear algebra (HHL, variational linear system solver) 
        \\
        F7 & Primitives (Grover\rq s algorithm, amplitude amplification,\\ 
        ~ & phase estimation, variational algorithm, optimizers,\\ 
        ~ & measurements)\\
        F8 & Simulating quantum mechanics 
        \\ 
        F9 & High energy physics
        \\ 
        F10 & Fluid dynamics and other differential equations 
        \\ 
        F11 & Open quantum systems 
        \\ 
        F12 & Quantum walk 
        \\ 
        F13 & Others 
        \\ \hline
    \end{tabular}
    \caption{(Left) List of the topics in our survey. Here topic A includes the Bell inequality\cite{PhysicsPhysiqueFizika.1.195}, entanglement test (Greenberger-Horne-Zeilinger (GHZ) state\cite{Greenberger1989}, graph state\cite{PhysRevLett.86.910}), and protocols including quantum teleportation\cite{PhysRevLett.70.1895}. Topic C contains  research about compilers, circuit optimization, gate decompositions, benchmarking different quantum computers, benchmarking methods such as cross-entropy benchmarking (XEB)\cite{Boixo2018-tc}, and QV\cite{PhysRevA.100.032328}.
    (Right) List of the subtopics in the topic F: Applications.
    QAOA and HHL stand for quantum approximate optimization algorithm\cite{farhi2014quantum} and Harrow-Hassidim-Lloyd algorithm\cite{PhysRevLett.103.150502}, respectively.}
    \label{topics}
 \end{table}

Under the above specification, we retrieved the manuscripts from the quant-ph section of arXiv in two ways. In the first way,
we retrieved the manuscripts whose abstracts contain one of the company names or AWS (Amazon Web Service). The latter keyword AWS was adopted by taking into account that QCs by IonQ or Rigetti are used through AWS. In the second way, we
retrieved the manuscripts that includes the leaders of QC research in the above companies among the authors. Here the leaders
are: Hartmut Neven for Google, Jay M. Gambetta for IBM, Christopher Monroe for IonQ, and Chad Rigetti for Rigetti. For
Quantinuum, through a preliminary investigation of the papers and website\cite{Quantinuum}, we identified Ross Duncan, Stephen Clark, Steve
Sanders, Patricia J. Lee, Marcello Benedetti, Chris Langer, Jenni Strabley, Russell Stutz, Brian Neyenhuis, and Henrik Dreyer
as the leaders possibly using the QCs. 
These two kinds of retrievals add up to 1204 preprints excluding the duplicates.

Given the set of the preprints automatically retrieved by using arXiv API, the authors of this review read the preprints and identified the preprints that employed actual QCs. 
We first excluded the preprints whose authors we could find with certainty that did not employ QCs just by reading the abstracts.
We next read the body of the remaining preprints to identify whether the authors indeed employed QCs of the five vendors (based on the gate model, not on the quantum annealing or quantum simulator) and extracted the following attributes: a) topics of the preprint (up to 2), b) presence of authors affiliated with the vendors, c) the names of the QPUs employed and their vendors (up to 3), and d) the maximal number of the employed qubits in the preprint.
In these elimination steps, we singled out 748 preprints.
The topics are designed by reading all the abstracts of the preprints.
See Table \ref{topics}.

Here we argue whether our dataset is large enough to grasp the present situation of the research and development of the QCs.
For this purpose, it is notable to mention that IBM Quantum Network members have published more than 675 academic publications in quantum information science \cite{IBM_pubs}.
On the other hand, our dataset has 614 preprints using the IBM\rq s QCs.
Our dataset covers approximately 91.0\% of the number of publications by IBM Quantum Network members.
Therefore, we can say that our dataset is large enough for the following analyses.

Prior to the presentation of our results, let us mention how we draw the box-and-whisker plots of our paper (see Fig. \ref{general} (c), for example).
Given the data and their attributes to be plotted, the length of the upper (lower) whisker is set as a maximum (minimum) value not exceeding 1.5 times the interquartile range (IQR, the height of the box). 
Data whose attribute values are outside this range are called outliers.
Throughout this paper, we do not plot the outliers in the box-and-whisker plots, in order to see the typical behaviors of the data in detail.
On the other hand, we hereafter use the terminology \lq\lq maximal\rq\rq~ and \lq\lq max\rq\rq~ by taking the outliers into account.


\section*{Results}

\subsection*{Annual changes and vendor share}

\begin{figure}[t]
\centering
\subfigure[]{
\includegraphics[width=.42\linewidth]{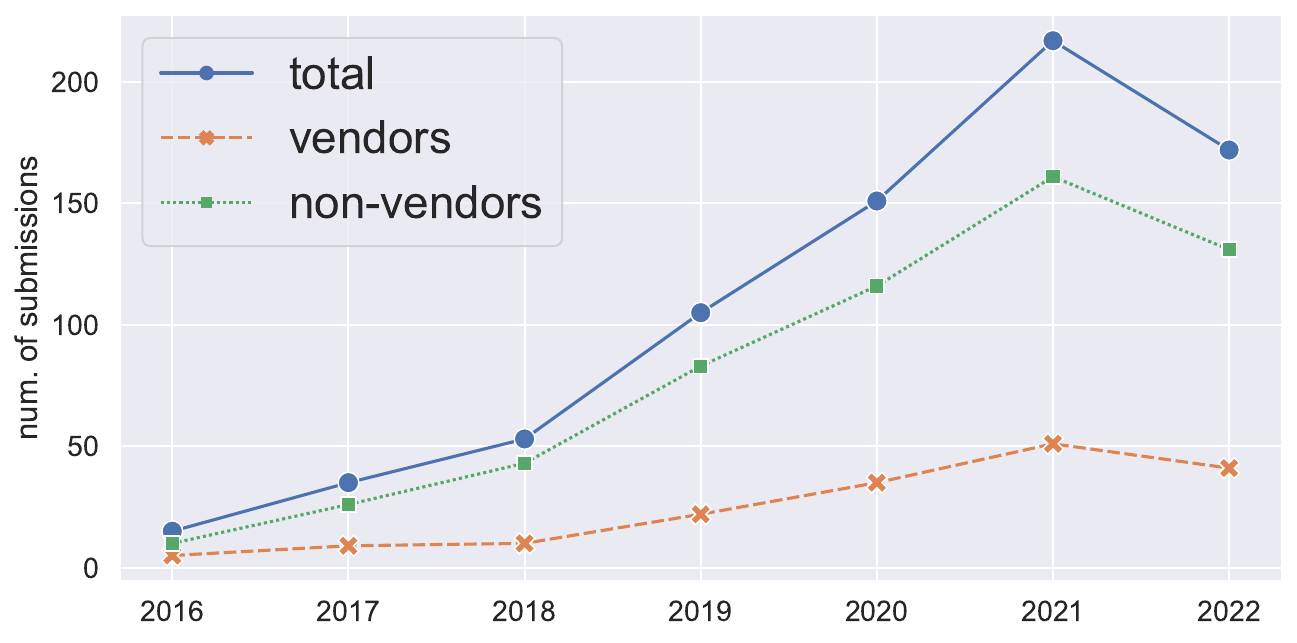}
}
\subfigure[]{
\includegraphics[width=.42\linewidth]{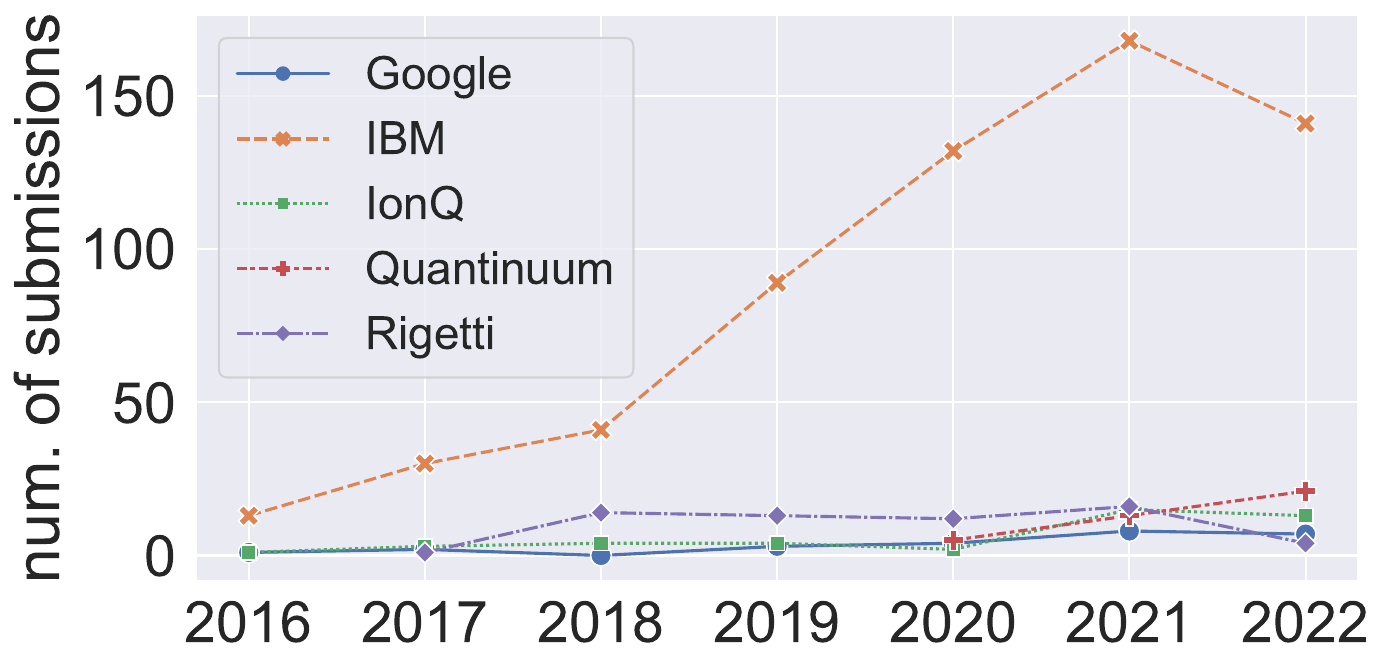}
}
\\
\subfigure[]{
\includegraphics[width=.42\linewidth]{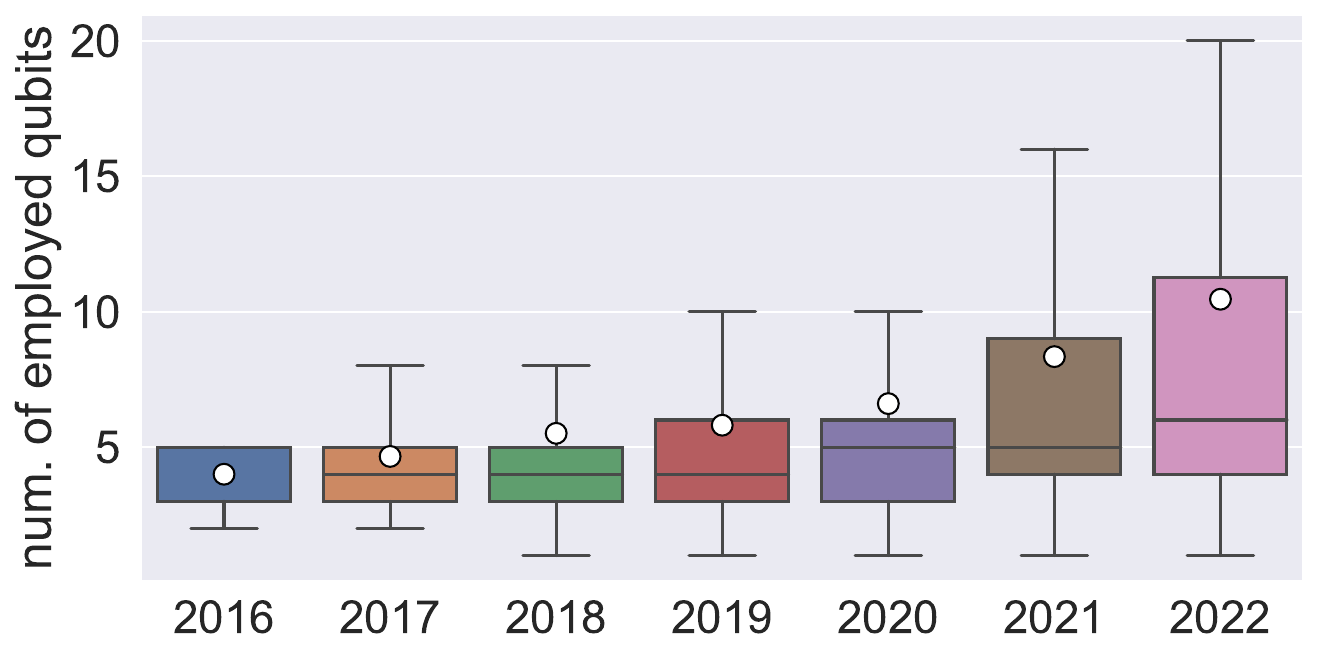}
}
\subfigure[]{
\includegraphics[width=.42\linewidth]{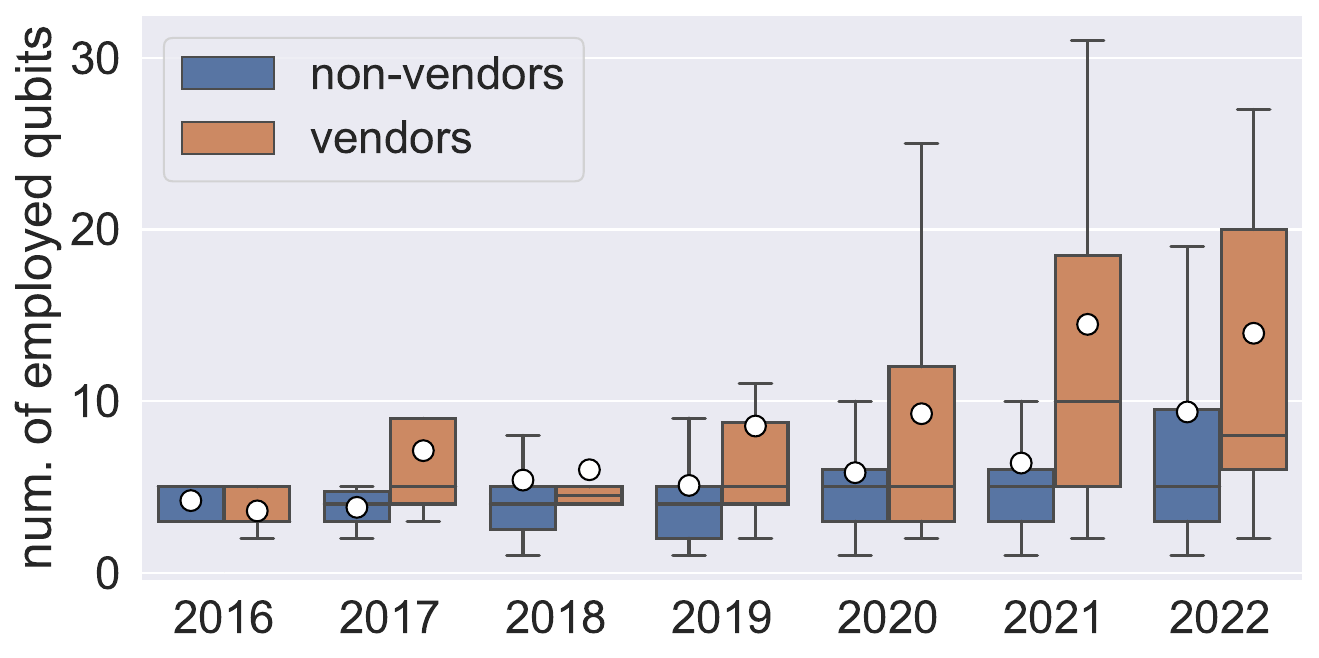}
}
\\
\subfigure[]{
\includegraphics[width=.42\linewidth]{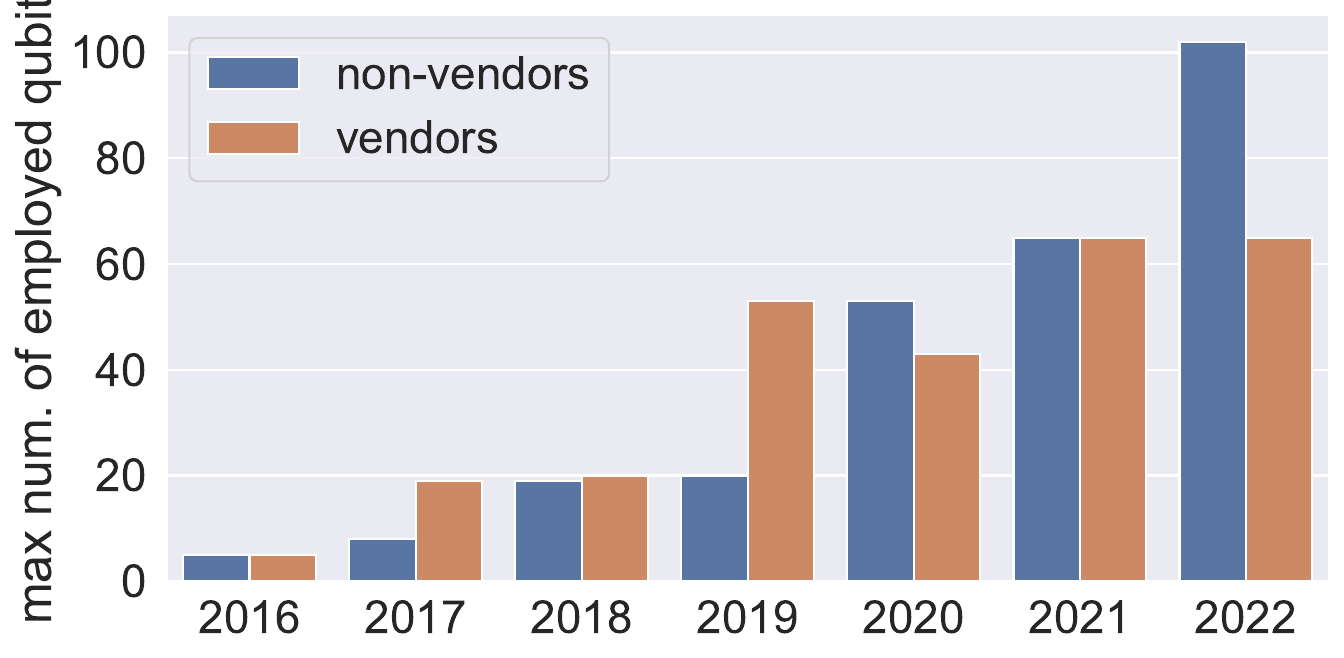}
}
\caption{(a, b) Number of submissions per year.
(c, d) The box-and-whisker plots without the outliers for the number of employed qubits per year.
The white circles show the averages.
(e) The maximal number of employed qubits per year.
Graphs (a, d, e) are stratified by the presence of the authors in the vendor companies, whereas (b) is by the vendors of the employed QPUs.
In our dataset, up to three vendor names can be written per preprint. 
In addition, these vendor names were treated independently when drawing graph (b). 
For these reasons, the total number of submissions calculated from graph (b) exceeds that calculated from the dataset.
}
\label{general}
\end{figure}

\begin{figure}[t]
\centering
\includegraphics[width=.55\linewidth]{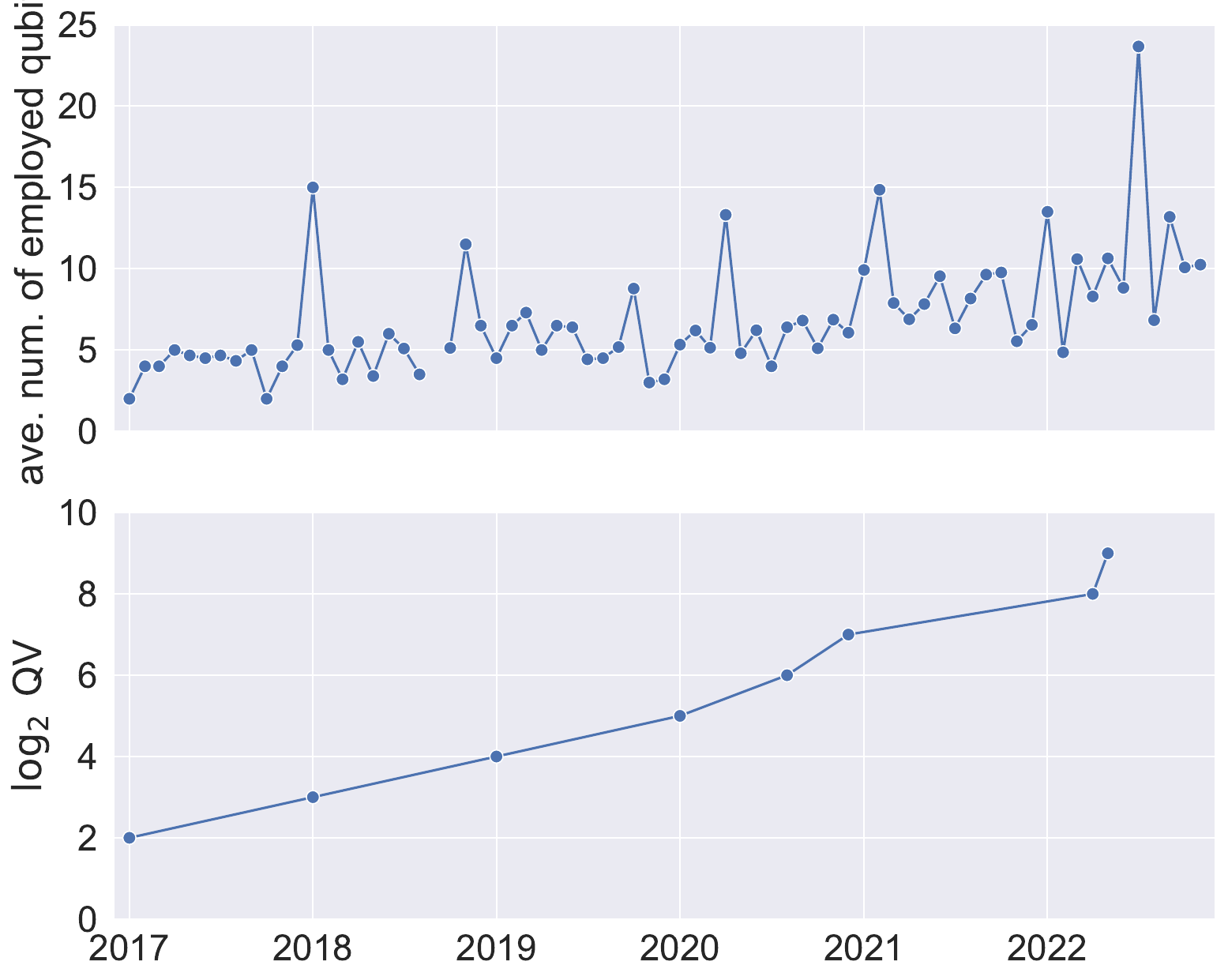}
\caption{
The monthly average number of the qubits employed (above) and the updates of the QV of IBM\rq s QPUs\cite{mediumWhatQuantum,forbesDoublesQuantum,zdnetHitsQuantum,Gambetta_2020,ibmQuantumAchieved,Gambetta_2022} (below). 
In the graph above, the average is missing for September 2018, because no preprints actually using QCs are retrieved for this month.
In the graph below, as for the year 2019 and before, the months when the values of QV were taken are set to January for convenience, since those months are unmentioned in the data source\cite{mediumWhatQuantum}.
}
\label{qv_cor}
\end{figure}

\begin{figure}[ht]
\centering
\subfigure[]{
\includegraphics[width=.32\linewidth]{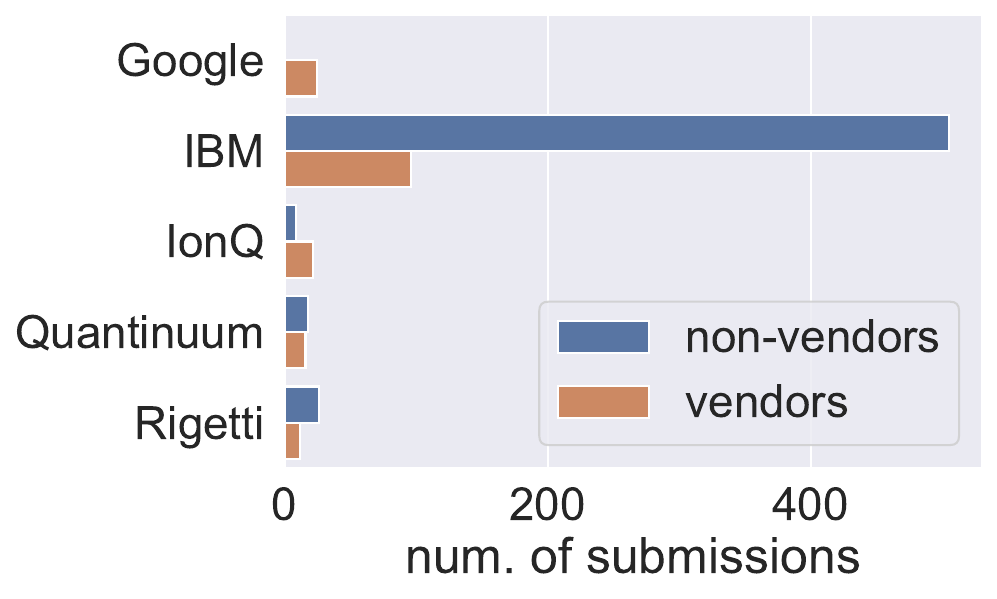}
}
\subfigure[]{
\includegraphics[width=.32\linewidth]{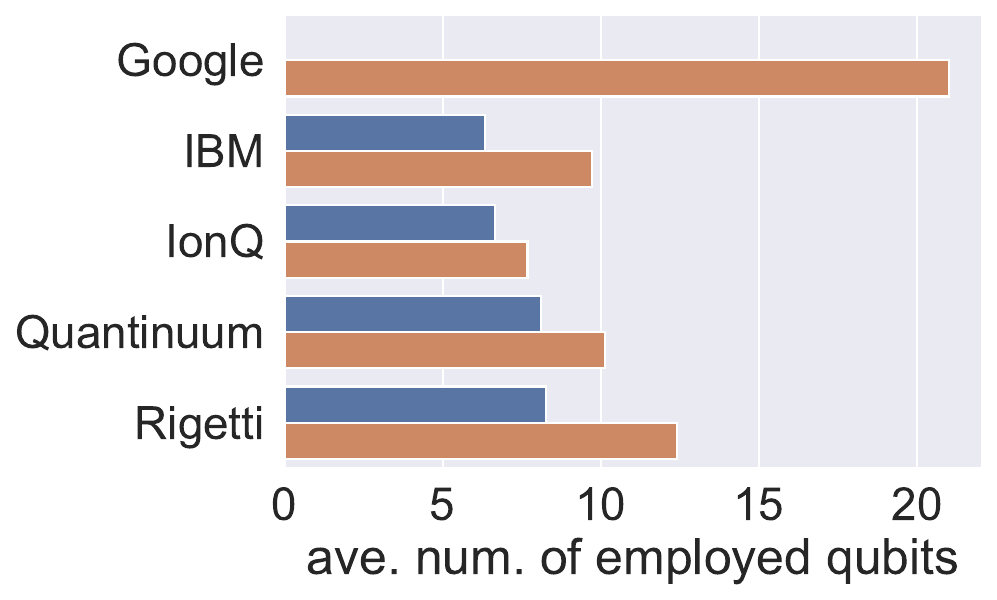}
}
\subfigure[]{
\includegraphics[width=.32\linewidth]{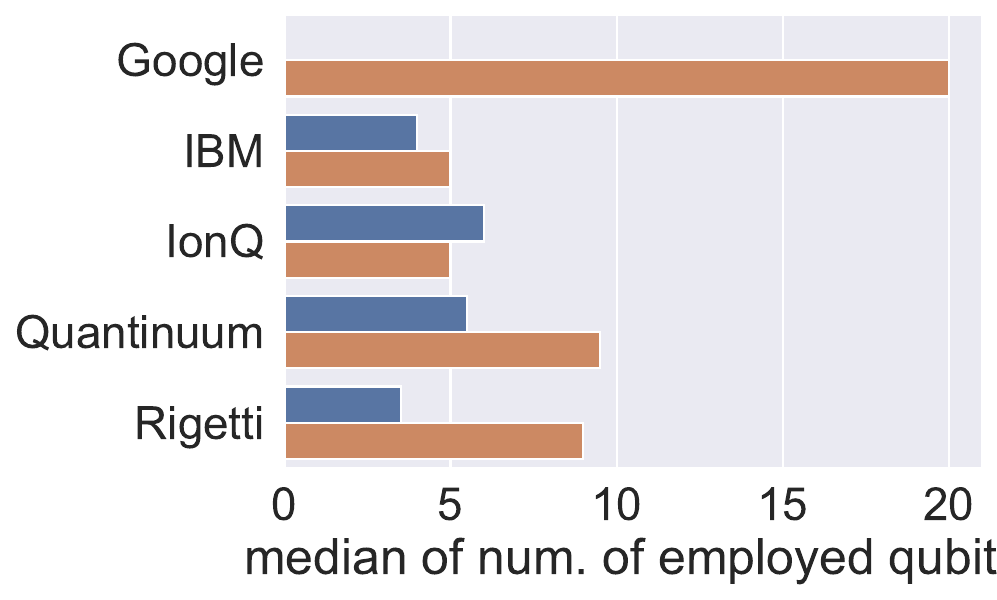}
}
\caption{
(a) Number of submissions for all years per the vendor of the QPU that the authors employed.
(b) Average of the number of employed qubits for all years per the vendor of the QPU that the authors employed.
(c) Median of the number of employed qubits per the vendor of the QPU that the authors employed.
All the graphs are stratified by the presence of the authors in the vendor companies.
}
\label{vendor_strategy}
\end{figure}


Figure \ref{general} (a) shows the general trends of the number of manuscripts using QCs.
The annual number of submissions increases up until 2021.
In particular, the increase in speed was accelerating since 2018: the slope from 2018 to 2021 is about 1.16 times larger than that from 2016 to 2018.
One of the reasons for this acceleration is the release of the programming language Qiskit on March 7 of 2017\cite{Gambetta2018}, because Ref.\citeonline{Gambetta2018} reports the sudden increase in the number of executions of IBM Quantum (formerly called IBM Q) after the release of Qiskit and most of the QPUs employed seen in our dataset around that time are of IBM (Fig. \ref{general} (b)).
Note that Fig. \ref{general} (b) also shows the annual change in the number of preprints using each vendor's QPUs.
We can find not only the IBM\rq s dominance in share over the observed years but also the rapid growth in Quantinuum\rq s share since 2020.
Indeed, the share of the number of submissions using each vendor\rq s QPUs in 2022 are 3.8\% for Google, 75.8\% for IBM, 7.0\% for IonQ, 11.3\% for Quantinuum, and 2.2\% for Rigetti, respectively. 
In contrast, the number of submissions decreases from 2022, mainly due to fewer submissions from outside the vendor companies (Fig. \ref{general} (a)).
However, this decrease may partially result from the fact that the target period of our survey ends on November 10 of 2022.
If we compensate for the influence of this lacking two months, then the number of the estimated submissions in 2022 is approximately 206.4 (=$12/10\times 172$, where 172 is the number of the submissions  in 2022 observed through this survey), which is comparable with the number of the submissions 217 in 2021.
 In either case, the slowdown of the growth rate in the number of submissions is evident.
Although it might indicate that the use of QCs has genuinely reached a plateau, the saturation could be an artifact resulting from our method of retrieving manuscripts, specifically targeting those that include the vendor's name in their abstracts. 
The authors may have become less inclined to write the vendor's name in the abstracts \cite{shirakawa2021automatic}, presumably due to the increasing popularity of utilizing IBM's actual machines.
In conclusion, the use of the QCs has become popular with the help of the programming language Qiskit, and now reaches a plateau, at least apparently.

\begin{figure}
\centering
\subfigure[]{
\includegraphics[width=.85\linewidth]{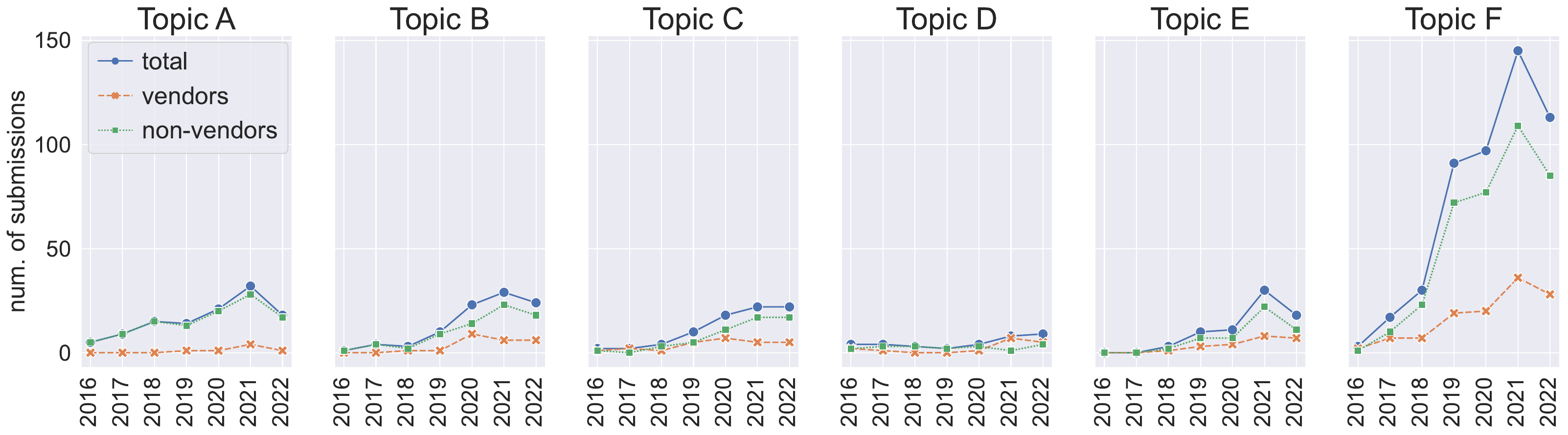}
}
\\
\subfigure[]{
\includegraphics[width=.85\linewidth]{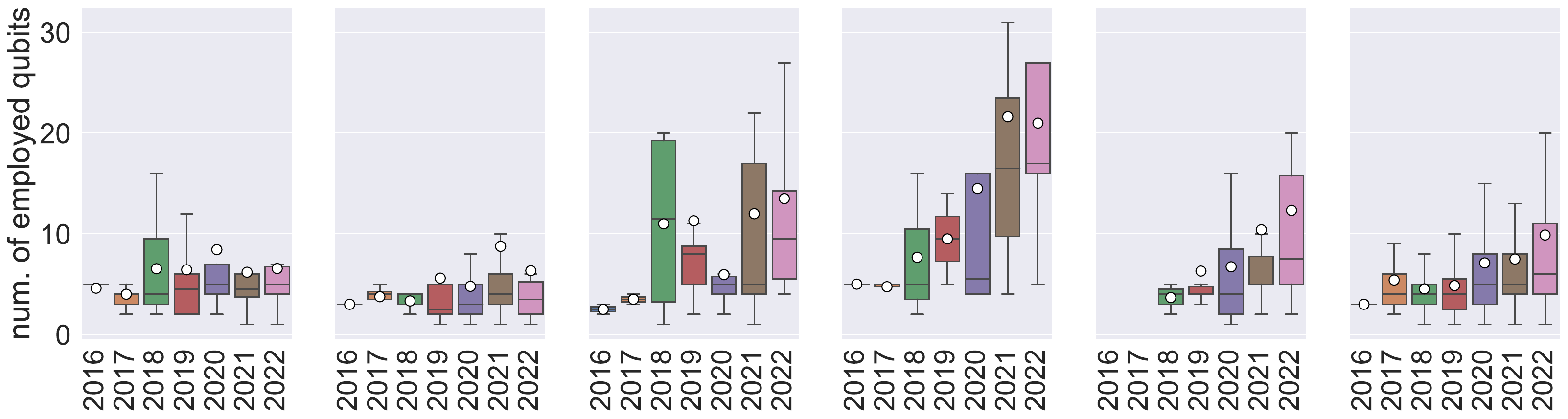}
}
\\
\subfigure[]{
\includegraphics[width=.85\linewidth]{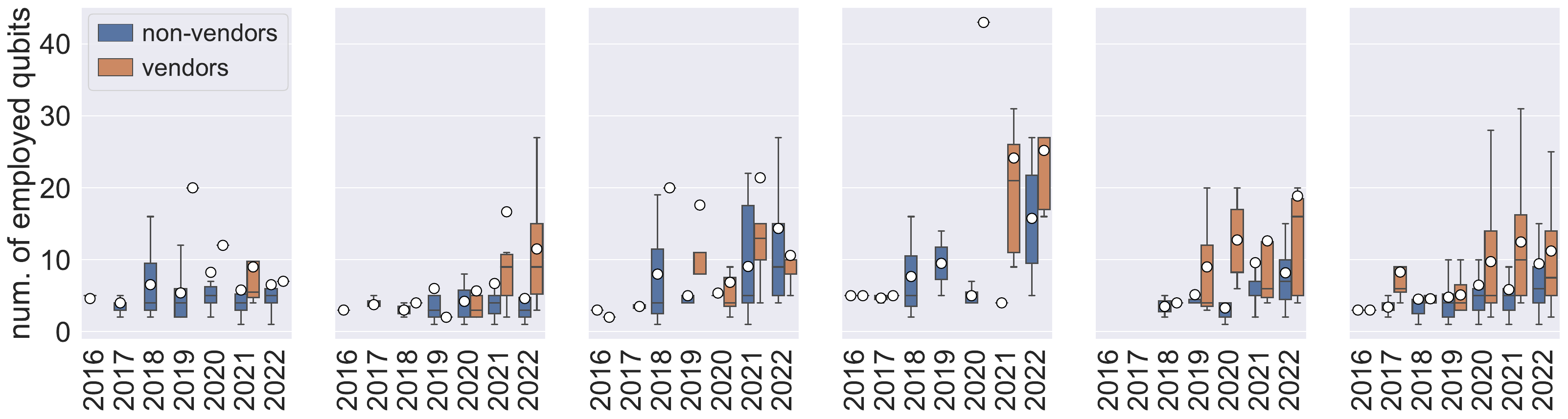}
}
\caption{(a) The topic-wise number of submissions per year. 
In our dataset, up to two topics can be written per preprint, and we treated these topics independently when drawing graph (a). 
Thus, same as Fig. \ref{general} (b), the total number of submissions calculated from graph (a) exceeds that calculated from the dataset.
(b, c) The topic-wise box-and-whisker plots without the outliers for the number of employed qubits per year. 
Graphs (a) and (c) are stratified by the presence of the authors affiliated with the vendor companies.
See Table \ref{topics} for the details of the topics.
}

\label{topicwise}
\end{figure}

Let us proceed to see the typical number of qubits employed.
The box-and-whisker plots in Fig. \ref{general} (c) for the number of the employed qubits show that the upper end of the whisker reaches 20 in 2022 whereas the lower end has been one except two in 2016 and 2017.
Since the size of the QPUs of IBM has scaled in accordance with the roadmap\cite{IBM2023}, the expansion of the upper whisker implies that some authors make good use of the larger QPUs.
Such authors are likely to be affiliated with the vendor companies, according to Fig. \ref{general} (d).

In contrast, the median of the number of qubits employed has been almost stable: 5 qubits in 2016 to 6 in 2022.
This is because the NISQ devices are not robust against noise and errors, and thereby the outcomes from the computation are, in general, not reliable if one employs a large number of qubits.
Indeed, we can find such a case in Ref.\citeonline{Lavor2022-ez}, where the authors restricted themselves to using 3 qubits in order to avoid the influence of noise and errors.
On the other hand, when examining Fig. \ref{general} (d), it becomes evident that preprints associated with vendor companies have contributed to an increase in the median number of employed qubits, reaching 8 in 2022.
As shown below, it can be observed that this increase in the median number of qubits is particularly noticeable in two areas: (D) quantum error correction\cite{McEwen2021-pl,Chen2021-at} and (E) quantum noise mitigation\cite{obrien2022purificationbased}. In these areas, the author affiliated with the vendors has recently been employing a large number of qubits, typically $\approx 20$ qubits.
One of the reasons for this increase could be that the vendors have their roadmaps (Ref. \citeonline{IBM2023}, for example) which announce an increase in the size of the QPUs and use of them in the near future.
This could provide an incentive for vendor researchers to experiment with a larger number of qubits, resulting in an increase in the median.

\begin{figure}[t]
\centering
\subfigure[]{
\includegraphics[width=.85\linewidth]{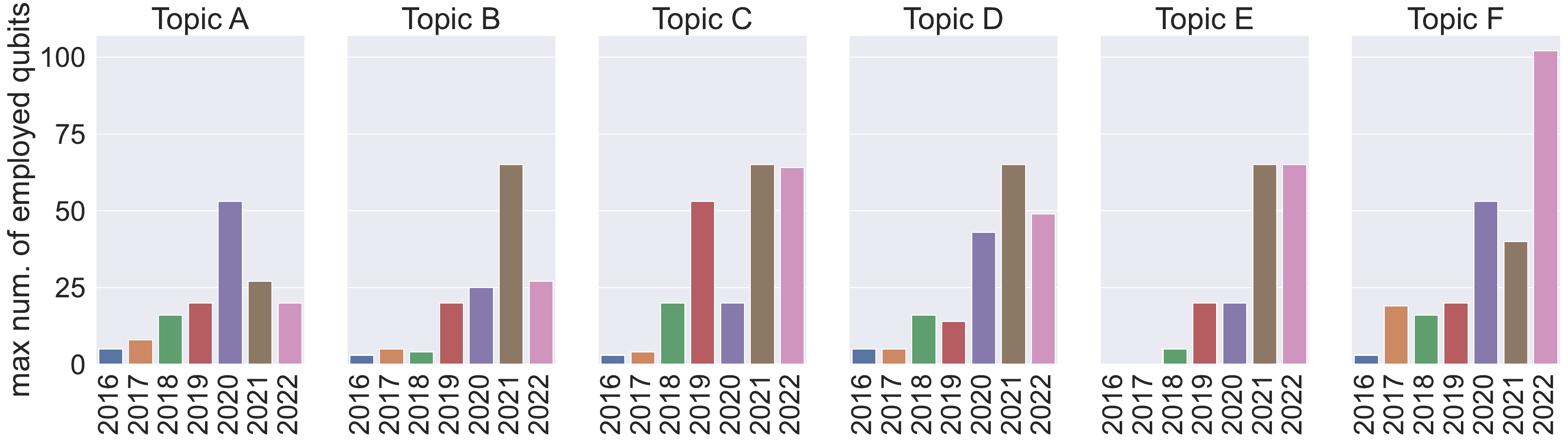}
}
\\
\subfigure[]{
\includegraphics[width=.85\linewidth]{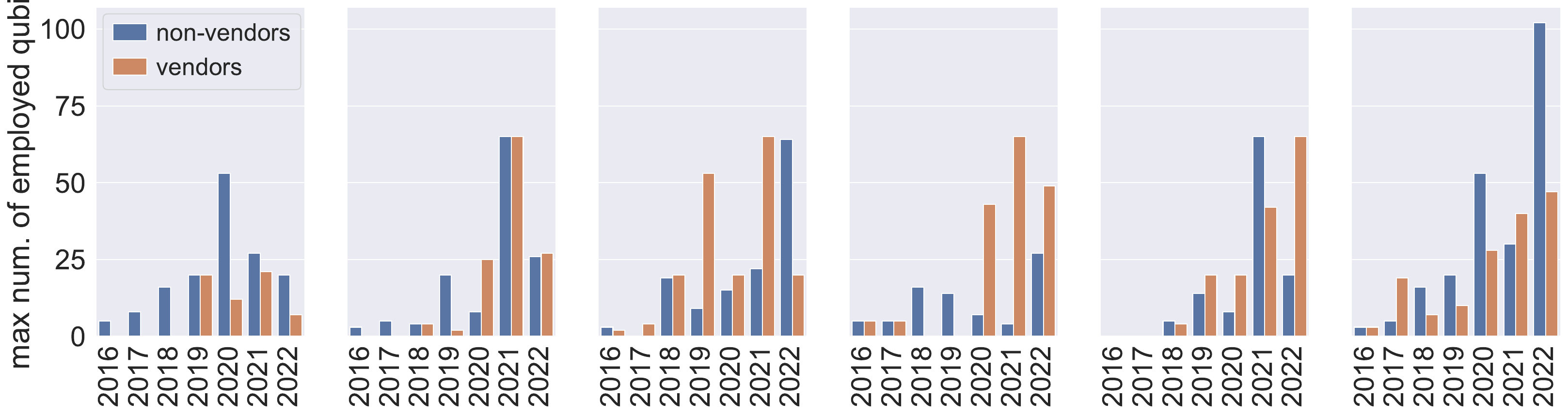}
}
\caption{(a) The topic-wise bar plots for the maximal number of employed qubits per year. (b) The topic-wise bar plots for the maximal number of employed qubits per year stratified by the presence of the authors affiliated with the vendor companies.
See Table \ref{topics} for the details of the topics.
}
\label{topicwise_max}
\end{figure}

The average number of employed qubits has increased almost 2.5 times from 4 in 2016 to 10.46 in 2022 (Fig. \ref{general}(c)).
Besides, the average is larger than the median over the years.
Since the average is sensitive to the existence of outliers, this difference between the median and average suggests that the outliers raise the averages considerably.
Indeed, the maximal number of employed qubits rapidly increases from 5 in 2016 to 102 in 2022 (Fig. \ref{general} (e)).
Note that the maximal number of qubits in each year is achieved mainly in four topics: (C) system and software development, (D) quantum error correction, (E) quantum noise mitigation, and (F) applications. 
More precisely, the maximum of each year is attained by applications on open quantum systems\cite{doi:10.1126/sciadv.1700672}, condensed matter physics\cite{Zhang2017,PhysRevResearch.2.043205,PhysRevResearch.5.013183}, the proposal of the quantum volume (QV)\cite{PhysRevA.100.032328}, quantum supremacy experiment\cite{Arute2019-rf}, quantum noise mitigation technique to generate whole-device entangled states\cite{https://doi.org/10.1002/qute.202100061}, hexagonal matching code\cite{wootton2022hexagonal}, to name a few.
Such good practices, including those appearing after this survey period, are investigated in detail in the subsequent subsection.

Now we turn to discuss a possible cause of the increase in the number of employed qubits. Figure \ref{qv_cor} shows that the monthly change in the average number of employed qubits correlates with the logarithm of the QV of the IBM devices since 2017: both have increased by about four times from 2017 to 2022.
This suggests that the QV properly captures the quality of qubits and it is also a good yardstick for the typical number of qubits we can manage to manipulate at least for the standard usages of the QCs, where the quantum circuits to be implemented are not necessarily shallow.

Figures \ref{vendor_strategy} (a), (b), and (c)  show the number of submissions, average number of the employed qubits, and median over the observed years respectively, classified in accordance with the vendors of the QPUs that the authors used.
From these figures, we may read the vendors\rq~present statuses and strategies:
IBM\rq s QPUs dominate in the number of submissions.
In contrast, Google restricts the use of its QPUs only to research groups with the participation of in-house researchers, who conduct experiments typically employing as many as 20 qubits.
The average number of employed qubits in the preprints using the QPUs of Quantinuum \cite{PhysRevA.101.010301,PhysRevApplied.16.044057,Amaro_2022,kikuchi2023realization} and Rigetti are larger than those of IBM and IonQ.
Note that, although the two outstanding companies (Google and IBM) in the analyses adopt the superconducting qubits, these figures are not to claim the superiority of superconducting devices; it is still unclear what kinds of physical systems are suitable for the large-scale implementation of the QCs \cite{NAP25196}.



\subsection*{Topics and applications}

\begin{figure}
\centering
\subfigure[]{
\includegraphics[width=.85\linewidth]{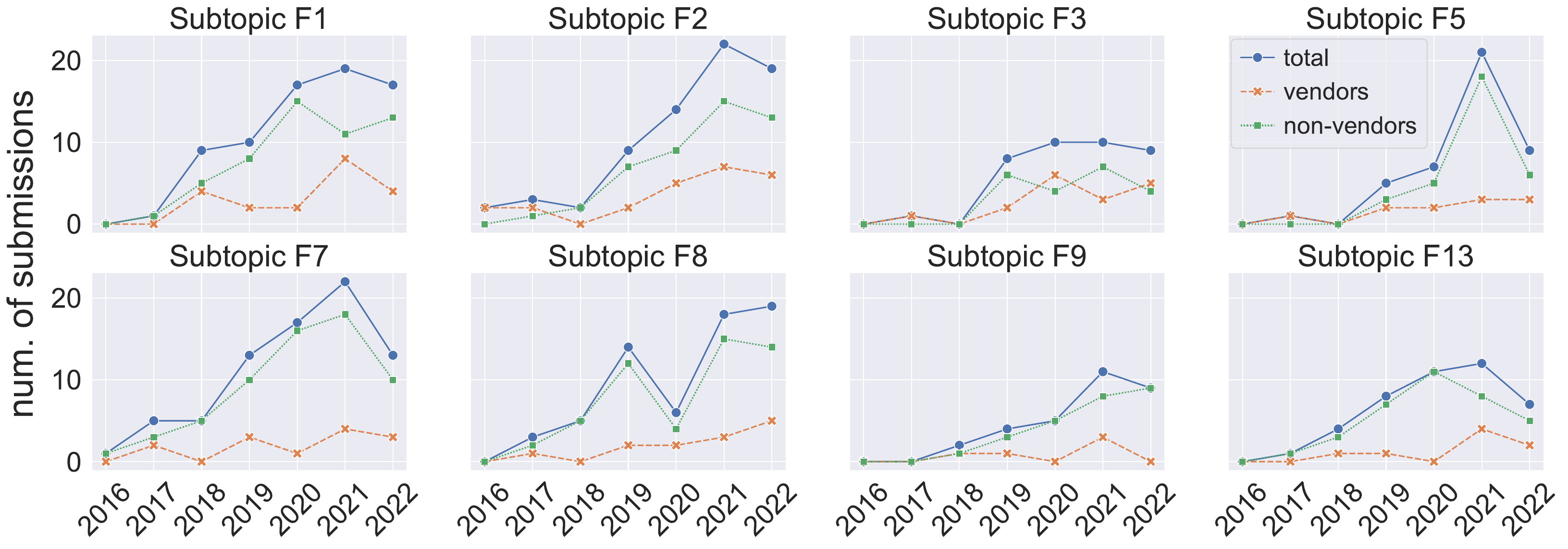}
}
\\
\subfigure[]{
\includegraphics[width=.85\linewidth]{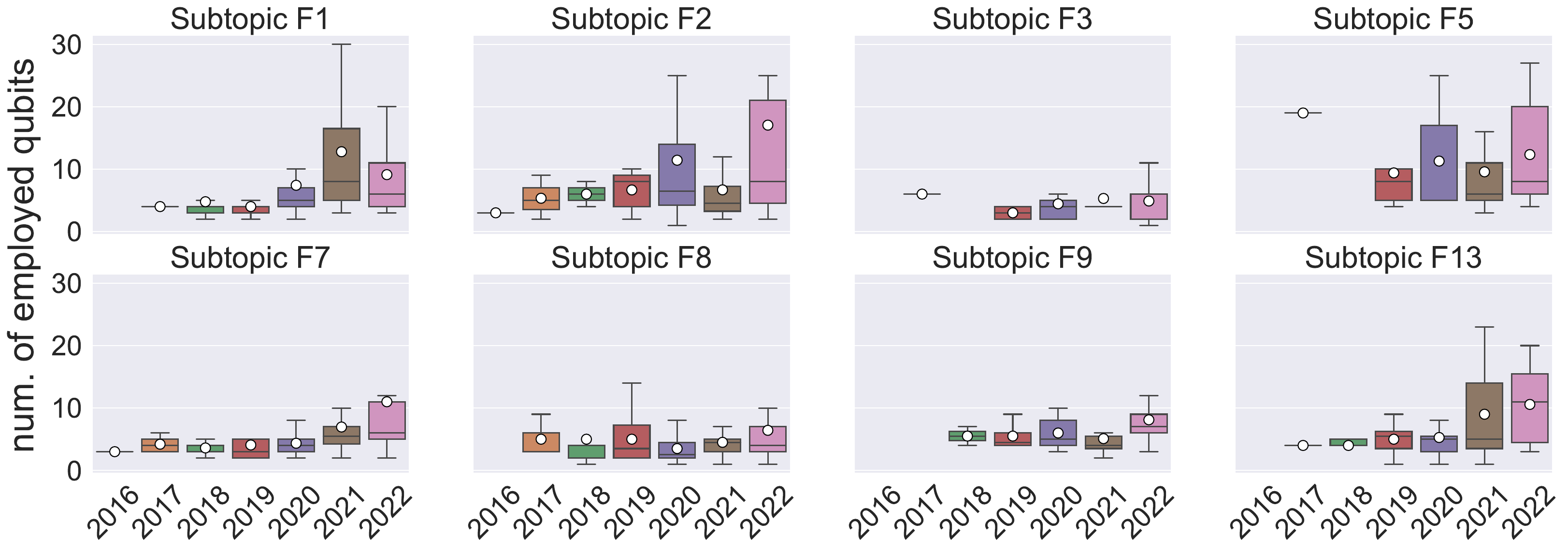}
}
\\
\subfigure[]{
\includegraphics[width=.85\linewidth]{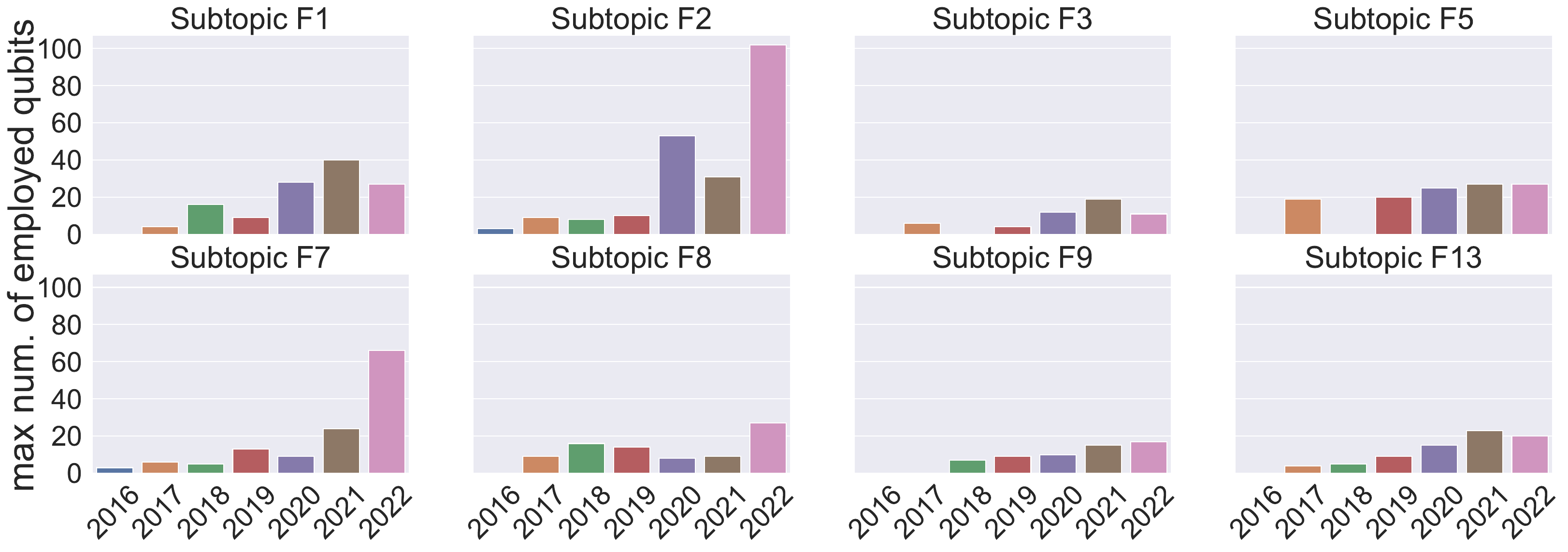}
}
\caption{Graphs for the preprints of the subtopics in the topic F: Applications. (a) The subtopic-wise number of submissions per year stratified by the presence of the authors affiliated with the vendor companies. (b) The subtopic-wise box-and-whisker plots without the outliers for the number of employed qubits per year. (c) The subtopic-wise maximal number of employed qubits per year.
See Table \ref{topics} for the details of the subtopics.
}
\label{subclass}
\end{figure}

Now we take a closer look at the topic-wise trends of the usages of the QCs.
For the number of submissions, Fig. \ref{topicwise} (a) shows that all but two topics, (C) system and software development and (D) quantum error correction, have trends similar to the overall one observed in Fig.~\ref{general} (a): a gradual increase until 2021 with acceleration from 2018, and a decrease in 2022.
In particular, the topic (F) applications accounts for more than 50\% of the submissions in 2022, and the speed of increase in the number of submissions has accelerated since 2018.
These facts imply that the majority of the users of the QCs have
 their interests mainly in the applications, and Qiskit has facilitated their use of the QCs.
On the other hand, topics C and D have the number of submissions in 2022 as many as that in 2021, showing the growing interest in these topics.

For the number of the employed qubits, we can observe distinct behaviors depending on the topics (Fig. \ref{topicwise} (b)): Topics A (Fundamentals of physics and quantum information), and B (Tomography, noise characterization, quantum control (including pulse
optimization), gate benchmarking), have approximately constant whisker lengths and medians, whereas the other topics have the expansion of the whiskers and the increasing medians, suggesting the scaling up of the experiments associated with the latter topics.
In particular, topic D (Quantum error correction) has experienced an outstanding increase in the number of employed qubits since 2018.
Furthermore, by introducing the stratification by the presence of the vendor authors, we can more precisely reveal what the authors employ a large number of the qubits for. 
Figure \ref{topicwise} (c) clearly shows that the authors in the vendor companies use many qubits for the experiments in topics B, D, and E (Quantum noise mitigation), while so do the non-vendor authors in topics C (System and software development) and F (Applications).
Note that the vendor authors\rq~ behavior to use a large number of the qubits for the above topics is consistent with the roadmap of a vendor company\cite{IBM2023} which aims to improve the accuracy of the quantum computation by implementing scalable noise mitigation (topic E) and quantum error correction (topic D) in the near future.

The above observation obtained from Fig. \ref{topicwise} is consistent with what we can find in the annual change of the maximal number of the employed qubits (Fig. \ref{topicwise_max}).
Except for the spikes observed (at topic A in 2020, B in 2021, C in 2019, and F in 2020), all the topics tend to increase in the maximal number of the employed qubits through the survey period.
In particular, we can observe the rapid increase of the maximal number of the employed qubits in topics C, D, E, and F.
Similarly to Fig. \ref{topicwise} (c), the authors in the vendor companies contribute to the increase in topics D and E, whereas so do the non-vendor authors in topics C and F (Fig. \ref{topicwise_max} (b)).

We proceed to find what kinds of applications are favored in the preprints using the QCs.
This analysis requires further classification of the preprints of topic F (Applications) into the subtopics given in Table \ref{topics}, resulting in the small number of preprints remaining in each subtopic.
For this, we hereafter do not make use of the stratification by the presence of the vendor\rq s authors, except for Fig. \ref{subclass} (a), which plots the annual changes in the number of submissions.

Of the 113 preprints submitted in 2022, 102 preprints (90.3\%) are in the top eight subtopics (F1: Quantum machine learning, F2: Condensed matter physics, F3: Quantum chemistry, F5: Optimization (QAOA), F7: Primitives, F8: Simulating quantum mechanics, and F9: High energy physics, F13: Others).
We hereafter call these eight subtopics the major subtopics, all of which tendo to increase in the annual number of submissions over the survey period (Fig. \ref{subclass} (a)).

Among the 8 major subtopics, now we focus on the top four subtopics (F1, F2, F7, and F8) in terms of the number of submissions in 2022.
Each subtopic has more than 12 submissions in 2022, which represents more than 10\% of the total number of submissions to topic F (Applications) in 2022.
In particular, the subtopic F8 (Simulating quantum mechanics) is notable for the increase in the submissions from the vendor\rq s authors even in 2022, in spite of the lack of two months in the target period.
These submissions to F8 from the vendors\rq~authors include a 20-qubit simulation of non-equilibrium phase transition by Quantinuum\cite{chertkov2022characterizing}.
On the other hand, we can observe the increase in the submissions from the non-vendor\rq s authors in the subtopic F9 (High energy physics) in 2022, which includes the simulation of the single-baryon $\beta$-decay process by using 17 qubits in a Quantinuum device\cite{PhysRevD.107.054513}.

Let us further characterize the major subtopics by comparing the medians and averages of the number of the employed qubits with those obtained from all the preprints considered (Fig. \ref{subclass} (b)).
The subtopics F2 (Condensed matter physics), F5 (Optimization), F9 (High energy physics), and F13 (Others) have a median larger than the overall median 6 in 2022 (c.f. Fig. \ref{general} (c)).
Similarly, the subtopics F2, F5, F7 (Primitives), and F13 have an average larger than the overall average of 10.46 in 2022 (c.f. Fig. \ref{general} (c)).
Therefore, the common subtopics F2, F5, and F13 tend to use more qubits than all the (sub)topics. 
The subtopic F2 (Condensed matter physics) employs more qubits than the subtopic F3 (Quantum chemistry), although both the subtopics resemble each other in the energy scale of the systems to be treated.
One of the reasons for this difference lies in the simplicity of the Hamiltonian. 
Condensed matter physics treats the simplified Hamiltonians such as the Ising model, whereas quantum chemistry treats more complex ones.
Given the limited connectivity of the present NISQ devices, the complex Hamiltonians in quantum chemistry become hard to implement, as their size becomes large. 

In parallel, the subtopic F5 (Optimizations (QAOA)) recently employs a large number of qubits.
This can be understood from the fact QAOA can be performed with diagonal Ising dynamics,
which can be implemented by CNOT gates and single-qubit $Z$-rotations, and single-qubit $X$-rotations.
Furthermore, the cost function can be evaluated from only the $Z$-basis measurements since Ising Hamiltonian contains only $Z$ operators.These features simplify the ansatz and evaluation of the cost function of QAOA considerably than variational approaches
in the other subtopics such as quantum chemistry.
This may reduce effects of noise and statistical error leading to the increase in the number of employed qubits.

We also note that some major subtopics have outliers raising the average number of the employed qubits considerably.
The subtopic F7 (Primitives) has such outliers, whereas F9 (High energy physics) does not.
For example, by sorting the raw data with the help of Fig. \ref{subclass} (c), we can find the 100-qubit simulation of the ground state of the quantum spin chain\cite{PhysRevResearch.5.013183} in F2 (Condensed matter physics), the 27-qubit demonstration of solving resource allocation problem\cite{a15070224} in F5 (Optimization), the 66-qubit demonstration of acceleration of variational quantum solver by using parallelism\cite{Mineh_2023} in F7 (Primitives), and the 20-qubit implementation of a password authentication schemes on a quantum computer\cite{9605295} in F13 (Others).
Note that Fig. \ref{subclass} (c) shows a long-term increase in the maximal number of the employed qubits in the major subtopics.
Other notable works achieving the maximum of each year are: the 40-qubit demonstration of quantum learning of experimental data \cite{doi:10.1126/science.abn7293} in 2021, the 30-qubit embedding of classical data for machine learning\cite{https://doi.org/10.1002/qute.202100140} in 2022, both of which are for F1 (Quantum machine learning), and the 53-qubit generation of the GHZ states as the state preparation for the simulation of the condensed matter physics\cite{PhysRevResearch.2.043205} in F2 (Condensed matter physics).


\subsection*{Good practices}

Finally, we would like to discuss good practices and trends in the use of QCs. 
As mentioned in Fig. \ref{qv_cor}, the trends in the average number of qubits used can generally be correlated with the trends in QV. 
This is presumably because it is necessary to run relatively deep circuits in general-purpose applications of QCs. 
On the other hand, the current noise level of QCs does not allow for deeper quantum circuits when considering the use of more than 50 qubits. 
The practices of utilizing a relatively large number of qubits are therefore limited to shallow quantum circuits and we can divide them into three main categories: sampling tasks, simulations of quantum many-body systems, and demonstrations of quantum error correction.

{\it Sampling task}: Quantum computational supremacy of QCs over classical computers in random quantum circuit sampling was reported in 2019~\cite{Arute2019-rf}. 
This is defined as the task of simulating sampling from a quantum circuit that performs a random computation \lq\lq quantumly\rq\rq, rather than the task of computing any meaningful value. 
However, simulation technique on classical computers has improved dramatically since then, and it has now been shown that simulations on the scale of 53 qubits can be performed in the same amount of physical time.
On the other hand, stronger quantum computational supremacy with more qubits has recently been claimed~\cite{wu2021strong,ZHU2022240}. 
In particular, in Ref.~\citeonline{morvan2023phase}, random quantum circuit sampling with 70 qubits is presented, and it is concluded that even with improved classical simulations and the state-of-the-art supercomputer, it would take 47 years. 
Classical simulation is making progress, but it will eventually reach its limits. 
The next challenge would be to obtain quantum \lq\lq advantage\rq\rq~ in meaningful tasks such as the one discussed next.

{\it Quantum many-body system simulation}: Ref.~\citeonline{PhysRevResearch.5.013183} performs ground-state calculations of the one-dimensional (1D) XXZ model, an exactly solvable quantum spin model, in a relatively shallow circuit. 
The ground state energy of the 102-qubit system is calculated with an error of a few percent. 
Recent work by the IBM group ~\cite{kim2023evidence} has moved into areas that are even more difficult to simulate classically without considering the characteristic features of the system implemented.
In the IBM work, a 127-qubit quantum computer was used to estimate the expectation value of a physical quantity under the discretized two-dimensional (2D) transverse Ising dynamics which is intractable by most of the standard classical simulations such as matrix product states.
In contrast, this model is found to be simulated classically if we exploit the connectivity among the qubits in the system or truncate the sum of the Pauli operators resulting from the quantum circuit followed by the measurement of the physical quantity. 
See, for example, Refs.~\citeonline{tindall2023efficient,kechedzhi2023effective,begušić2023fast}.
It should be noted that, with the usage of quantum noise mitigation, both of the experiments indeed identified the expectation value of a physical quantity of interest to physicists, in a programmable way for large-scale quantum systems which are no longer possible to be simulated by brute-force classical computation.
The observation of dynamical phase transitions by a programmable quantum simulator with 53 qubits by IonQ~\cite{zhang2017observation}, Google's work on the generation of topologically ordered states using 31 qubits~\cite{satzinger2021realizing}, and the observation of the Majorana edge mode using 47 qubits~\cite{mi2022noise} are also in this direction.

{\it Quantum error correction}: From a different perspective than computational advantage, experiments using many qubits are demonstrations of quantum error correction, an essential element in the realization of large-scale QCs. 
In Ref.~\citeonline{wootton2020benchmarking}, 43 qubits were implemented using Qiskit-Ignis RepetitionCode and GraphDecoder to estimate the logical error probability to a code distance of 22. 
More recently, the Google group has conducted a demonstration of quantum error correction of surface codes using 49 qubits and reported that the logical error probability is reduced by increasing the code distance from 3 to 5~\cite{google2023suppressing}. 
As the number of qubits and error probability are entering the error-correctable range, we expect to see more reports of quantum error correction experiments in the future.

To summarize, while only small number of qubits corresponding to QV are employed for general-purpose applications, experiments using a relatively large number of qubits are being conducted in the direction of sampling tasks, quantum many-body simulation that reflect qubit topology, and quantum error correction experiments. These would be a near-term frontier of use cases of multi-qubit quantum computers beyond classical simulation.

\section*{Conclusion}

In this paper, we have conducted a quantitative survey on the recent trends in research using the QCs.
The important lesson we can learn from this survey is that the research using the actual QCs has matured enough at least in the number of publications. 
Many of these submissions are written by non-vendor authors and are devoted to the application of the QCs to various research fields.
More precisely, much literature has been devoted to expanding the scope of the application of the QCs to machine learning, condensed matter physics, quantum chemistry, optimization problems, primitives, simulations of quantum systems, and high energy physics.
The typical number of employed qubits in such works is growing along with the increase in QV.
This correlation implies that the majority of the QC users could employ relatively deeper circuits if we see this correlation by taking into account a trade-off between the number of the qubits reliably employed and the depth of the quantum circuits. 
In particular, the typical number of employed qubits is rapidly growing in the demonstrations of quantum error correction.

Moreover, we have found that the typical number of employed qubits is about six to ten, far from the maximum number of qubits available in the latest QPUs.
One of the reasons for this gap is the inevitable accumulation of noise and error effects during the computation due to the lack of implementation of the quantum error correction in the present NISQ devices.
As shown in our presentation of the good practices, the tasks performed in the experiments exceptionally utilizing a large number of qubits are the demonstrations of quantum computational supremacy, quantum many-body simulations, and the demonstrations of quantum error correction, all of which are implemented in shallow quantum circuits.

For the vendor share, we found that the number of submissions using IBM devices dominates and that using Quantinuum devices is rapidly increasing.
Google employs more qubits than the others, whereas Quantinuum and Rigetti follow Google in the typical number of employed qubits.

Although the targets of this survey are limited to the representative QC vendors, from a bird\rq s eye view, this survey has revealed interesting trends. 
One is the long-term temporal correlation between the average number of the employed qubits and the QV, suggesting that qualitative improvement of the NISQ architecture may benefit the use of more qubits.
The other is the vendor\rq s growing interest in noise characterization, gate benchmarking, quantum control schemes, quantum error correction, and quantum noise mitigation.
Indeed, the number of employed qubits is rapidly increasing in these topics, especially in the works by the vendor authors, although these topics require advanced techniques in the design and implementation of pulses and circuits.

On the basis of these observations, we can obtain the following outlook:
Quantum many-body system simulations of condensed matter physics and demonstrations of quantum error correction are shown to be feasible even in shallow circuits implemented in the existing NISQ devices, and we may expect more experiments on these topics with more qubits employed.
In parallel, we need to increase QV if we wish the increase the number of qubits employed by the non-vendor users who use QCs aiming at applications.
Both directions should be desirable to the sound development of quantum computation.

\bibliography{sample}


\section*{Acknowledgements}
TI thanks H. Okamuro for useful discussions. KF thanks Rie Fujii for helping data formatting.
This work was supported by MEXT Quantum Leap Flagship Program (MEXT Q-LEAP) Grant No. JPMXS0120319794 and JST COI-NEXT Grant Number JPMJPF2014. 
KMiya is supported by JSPS KAKENHI Grant No. JP22K11924.
YY is supported by JSPS KAKENHI Grant No. JP21K20536.
NT is supported by JSPS KAKENHI Grant No. JP19H05817, and JP19H05820.
HU is supported by JSPS KAKENHI Grant No. JP21H05182, JP21H05191, and 21H04446. 
HU and TS are supported by the COE research grant in computational science from Hyogo Prefecture and Kobe City through Foundation for Computational Science. 

\section*{Author contributions}
TI and KF designed the survey.
TI retrieved the preprints from the arXiv and KF designed the list of topics and subtopics.
All the authors extracted the attributes of the preprints.
TI made the analyses and wrote the manuscript through the discussion with all the authors.

\section*{Competing interests}
The authors declare no competing interests. 

\end{document}